# Relative efficiency of polariton emission in two-dimensional materials

*Siyuan Dai [†], Qiong Ma [‡], Yafang Yang [‡], Jeremy Rosenfeld [†], Michael D. Goldflam [†], Alex McLeod [†], Zhiyuan Sun [†], Trond I. Andersen [‡], Zhe Fei [†], Mengkun Liu [†], Yinming Shao [†], Kenji Watanabe [§], Takashi Taniguchi [§], Mark Thiemens [∥], Fritz Keilmann [□], Pablo Jarillo-Herrero [‡], Michael M. Fogler [†], D. N. Basov [†Δ*].*

[†] Department of Physics, University of California, San Diego, La Jolla, California 92093, USA

[‡] Department of Physics, Massachusetts Institute of Technology, Cambridge, Massachusetts 02215, USA

[§] National Institute for Materials Science, Namiki 1-1, Tsukuba, Ibaraki 305-0044, Japan

[∥] Department of Chemistry and Biochemistry, University of California, San Diego, La Jolla, California 92093, USA

[□] Ludwig-Maximilians-Universität and Center for Nanoscience, 80539 München, Germany

[Δ] Department of Physics, Columbia University, New York, New York 10027, USA

[*] Correspondence to: db3056@columbia.edu

**ABSTRACT:**

**We investigated emission and propagation of polaritons in a two dimensional van der Waals material hexagonal boron nitride (hBN). Our specific emphasis in this work is on hyperbolic phonon polariton emission that we investigated by means of scattering-type scanning near-field optical microscopy. Real-space nano-images detail how the polaritons are launched in several common arrangements including: light scattering by the edges of the crystal, metallic nanostructures deposited on the surface of hBN crystals, as well as random defects and impurities. Notably, the scanned tip of the near-field microscope is itself an efficient polariton launcher. Our analysis reveals that the scanning tips are superior to other types of emitters we have investigated. Furthermore, the study of polariton emission and emission efficiency may provide insights for development of polaritonic devices and for fundamental studies of collective modes in other van der Waals materials.**

Van der Waals (vdW) materials[1] have emerged as new media for the exploration of polaritons, the coupled collective oscillations of field and polarization charges[2-4]. Recently studied examples include *i*) plasmon-polaritons in conductors such as graphene[5-12], thin films of high-$T_c$ superconductors[13] and surface states of topological insulators (TIs)[14], *ii*) phonon-polaritons in insulators such as hexagonal boron nitride (hBN)[15-22] and bismuth-based TIs[14, 23], *iii*) hybrid plasmon-phonon-polariton modes in vdW heterostructures[14, 24-27]. The optical permittivity of vdW materials can be extremely anisotropic, having the opposite signs along the in- and out-of-plane axes, the property known as the hyperbolic response[28-31]. Accordingly, collective modes found in these frequency ranges are referred to as the hyperbolic polaritons. Appealing characteristics of polaritons in vdW systems include high optical confinement[5, 26, 32], low damping[32] as well as exceptional mechanical, optical, and electrical tunability[3, 7, 8, 21, 24, 25, 32, 33].

Coupling incident light to polaritons requires overcoming their momentum mismatch[2, 3]. This is possible if the system under study lacks translational invariance, for example, if the sample is of small size[15, 22], has a periodic patterning[5, 9, 11], or an intrinsic inhomogeneity. Alternatively, light-polariton conversion can be facilitated by extrinsic structures brought into the sample's proximity, such as metallic stripes or plates. Such methods have been utilized for emission of plasmon polaritons in graphene[34]. Emission of phonon polaritons by metallic edges have been demonstrated in SiC[35] and more recently in a vdW insulator hBN[16]. Below we report on these and other types of phonon-polariton emitters in hBN and evaluate their relative efficiency. Our analysis offers guidance for the development of nano-optical systems with high-efficiency coupling to polaritonic waves.

We focus on the mid-infrared (IR) spectral range $\omega = 1370 - 1610$ cm$^{-1}$ where hBN is optically hyperbolic[28-31]. Here $\omega = 1 / \lambda_{IR}$ and $\lambda_{IR}$ is the free-space IR wavelength. The polaritons existing in this domain are referred to as the hyperbolic phonon-polaritons (HP$^2$s)[16, 19, 21, 22, 25]. Real-space imaging of polaritons has been carried out using the scattering-type scanning near-field optical microscopy (s-SNOM, methods)[7, 8, 10, 16-21, 25, 32]. As shown schematically in Fig. 1a, under the illumination of an IR laser, the polaritonic standing waves can form and they can be visualized by scanning the sample under the tip of the atomic force microscope (AFM)[7, 8, 21]. A representative s-SNOM image (Fig. 1b) displays an oscillating pattern (or fringes) parallel to the crystal edges in a tapered hBN micro-crystal. The line profile (s-SNOM amplitude s($\omega$) as a function of the position $L$ along the blue line) associated with these fringes is plotted in Fig. 1c. Close to the hBN edges (white dashed lines), the s-SNOM image exhibits the strongest fringe followed by several other peaks with gradually decreasing amplitude (Fig. 1b-c). As established in previous studies[17, 20, 21], propagating HP$^2$s waves are both emitted and detected by the s-SNOM tip acting as an optical antenna[36]. These fringe patterns originate from interferences between the tip-emitted (solid purple arrow) and hBN edge-reflected HP$^2$s (dashed purple arrow). The measured fringe periodicity is equal to one-half of the polariton in-plane wavelength ($\lambda_p/2$)[16, 17, 20, 21] for the principal HP$^2$ branch. Interestingly, further away from the hBN edges, the $\lambda_p/2$ fringes become damped yet fringes with a much longer period persist (Fig. 1b-c). These latter fringes show the period of $\sim \lambda_p$ clearly visible in the line profile (Fig. 1c) and its Fourier transform (FT) (Fig. 1c inset, bottom).

We attribute the origin of the period-$\lambda_p$ fringes to polaritons emitted by the edges of hBN crystals marked with white dashed lines in Fig. 1b. The edge emission is illustrated by the annotated numerical simulation shown in the inset of Fig. 1a. The green arrows in this figure represent the monochromatic IR beam, which illuminates both the hBN crystal and the s-SNOM tip. This IR beam carries the momentum $k_{IR} = 2\pi / \lambda_{IR}$ arriving at the incidence angle $\psi$. At the hBN edge, the IR beam excites the HP$^2$ wave (cyan arrow) possessing the in-plane momentum of $k_p = 2\pi / \lambda_p$. This edge-launched wave propagates away from the edge and interferes with the incident IR beam (green arrows) forming a standing wave parallel to the edge. The measured s-SNOM signal is in the first approximation proportional to the local electric field underneath the tip apex[35]. This signal is produced by the field of the IR beam $E_i$ and that of the edge-emitted polaritons $E_p$: $E_{tip} = E_i \cos\psi + E_p$. The period of the corresponding interference fringes is[35] $\delta = 2\pi / (k_p - k_{IR} \cos\psi)$. In the case of highly-confined HP$^2$s in hBN[21, 22], $k_p$ is much larger than $k_{IR}$, and so $\delta$ is close to $2\pi / k_p = \lambda_p$, in accord with our data (Figs. 1b-c) and numerical simulations (Fig. 1a, inset). Clearly, when both edge-emitted and tip-emitted polaritons are present, the s-SNOM image may be rather complex. The observed pattern in Fig. 1b and 1c is the case in point.

The different amplitudes of the $\lambda_p$ and $\lambda_p/2$ polariton fringes seen in Figs. 1b and 1d (especially in the interior of hBN) are due to their distinct propagation trajectories and travel distances (Fig. 1a). Indeed, edge-emitted polaritons (fringe periodicity $\lambda_p$) propagate as plane waves, whereas tip-emitted polaritons ($\lambda_p/2$) propagate as circular ones. The amplitude of a circular wave decreases with the travel distance even in the absence of damping, whereas that of the plane wave does not. Furthermore, the path lengths of polaritons forming these $\lambda_p$ and $\lambda_p/2$ fringes are different. The edge-emitted polaritons ($\lambda_p$) only need to traverse the tip-edge distance $L$ once to become registered in our apparatus. The tip-emitted/edge-reflected polaritons (period $\lambda_p/2$) have to do a round trip. Therefore, the combination of geometric spreading and two-fold travel distance account for the faster decrease of period-$\lambda_p/2$ oscillations in Figs. 1b-c. In the interior of the hBN crystal (Fig. 1b), $\lambda_p$-periodic fringes emitted by the α edge (blue solid line in Fig. 1c) and β edge (red dashed line in Fig. 1c) exhibit different intensities. This difference is related to the shadowing of the sample by the AFM cantilever in our experiment. Once the hBN crystal is rotated by $\pi/2$, we observed identical $\lambda_p$ fringe intensity (Fig. 1d) from α and β edges, see Section 1 of the Supplementary Materials for details.

In addition to the hBN crystal edge and s-SNOM tip, polaritons in hBN are also emitted by metallic nanostructures. The images in Figs. 2a and 2c have been obtained from an hBN crystal with Au disks of height 188 nm and diameters 1-2 μm fabricated on its top surface. The s-SNOM data in Fig. 2a exhibit two groups of interference fringes. The fringes in the first group are parallel to the edge; those of the other group are concentric to the Au disk. To estimate the corresponding polariton wavelengths we examine the representative line traces (green and blue curves in Fig. 2b) extracted from the s-SNOM image in Fig. 2a by taking linear cuts along the lines of the same color. The green curve extracted from a line scan taken perpendicular to the hBN edge (Fig. 2a) again reveals a superposition of fringes with $\lambda_p/2$ and $\lambda_p$ periodicities. One can separate these components via the FT analysis[19]. In addition to the tip-emitted ($\lambda_p/2$) and edge-emitted ($\lambda_p$) polariton fringes (Fig. 2a) parallel to the hBN edges, one can witness concentric circular fringes around the Au disks (Fig. 2a and 2c). The profile for the concentric fringes arising from the gold emitter (blue curve, Fig. 2b) exhibits oscillations with nearly the same period as the edge emitted ones (black curve). Thus, it is attributed to polaritons emitted by the disk edges (magenta dashed circles,

Fig. 2a). To verify this assertion, we numerically simulated the s-SNOM image following an earlier study[35] (Fig. 3d); these simulations account for our experimental results.

Let us discuss the efficiency – the ability to convert incident IR photons into propagating polaritons – of different polariton emitters. We can quantify this parameter by taking the ratio of either the power $P_p$ or the intensity $I_p$ of the polariton wave to the incident IR power $P_0$ per unit area. The intensity scales as a square $I_p = |S|^2/M$ of the amplitude $S$ measured by s-SNOM. The proportionality coefficient $M$ is unknown but constant for a data set obtained for the same experimental conditions (IR frequency, illumination intensity, same tip, etc.). The ratio $l = I_p/P_0$, which has the units of length, defines the scattering length. This parameter is appropriate for an extended line-like emitter, such as an hBN edge. However, it is not suitable for a small, point-like emitter. As discussed above, such a point emitter generates waves whose intensity would decrease with distance $r$ (measured from its center) even in the absence of damping. In the latter case, it is the total power $P_p = 2\pi r I_p(r)$ of the wave that remains constant. The proper measure of efficiency is therefore the ratio $A = P_p/P_0$, the scattering cross-section, which has the units of area. The formula for the s-SNOM amplitude, which applies to both types of emitters and includes the phase of the wave and also unavoidable damping can be written as

$$S(r) = \sqrt{P_0 M}\, G(r) e^{i(q_p r + \phi)}, \qquad (1)$$

where $q_p$ and $\phi$ are the complex momentum and a phase shift, respectively. The "geometric" factor is $G(r) = \sqrt{l}$ for a line-like emitter and $G(r) = \sqrt{A/2\pi r}$ for a point-like one. Note that a metallic disk of radius $r_{disk} \gg \lambda_p$ is a type of emitter that can be considered a line-like near its edge, at $r - r_{disk} \ll r_{disk}$, and a point-like at $r \gg r_{disk}$. Accordingly, having determined its cross-section, one can also calculate the scattering length of the disk edge via $l_{disk} = A/2\pi r_{disk}$. For polaritons launched by the tip located a distance $x$ from the edge, one should substitute $2x$, the separation between the tip and its image[7, 8, 21] upon reflection, for $r$ in Eq. (1).

We have determined $A$ and $l$ by fitting the s-SNOM traces in Fig. 2b to Eq. (1). The fits are shown by the black dashed curves. Note that all of them have the same polariton damping factor $\gamma$ = Im $q_\mathrm{p}$ / Re $q_\mathrm{p}$ = 0.055 as in our previous studies[19, 21]. For the sake of comparing *relative* efficiency, we choose μm to be the unit of length, omit the common factor $\sqrt{P_0 M}$ in Eq. (1), and normalize $S(r)$ in suitable arbitrary units. From thus defined fitting procedure, we have obtained $A$ = 61 for the s-SNOM tip, $l$ = 0.56 for the hBN edge, $A$ = 14 and $l$ = 1.4 for the 1.5μm-radius Au disk (Table 1). We conclude that in terms of the polariton emission efficiency, the tip is superior to the disk treated as a point-like emitter; while as a line-like emitter disk is more efficient than the edge. We attribute the higher emission efficiency of the metallic objects (the tip and the disk) to their ability to concentrate electric field[36]. The hBN edge lacks this ability, which explains its comparatively low photon to polariton conversion efficiency. The higher emission efficiency of the tip compared to the disk indicates the former has a stronger coupling to high-$q_\mathrm{p}$ polaritons[7, 8, 15-17, 21]. Note that one possible way to calibrate the scattering parameters $A$ and $l$ in absolute units is by comparison to some standard emitters for which reliable theoretical calculations are possible. The hBN edge could be a candidate for such a standard emitter provided a better understanding of the field singularities at the sharp corners is developed, see Fig. 1a (inset) and Methods. This can be a subject of future work.

Impurities and defects on the sample surface (introduced unintentionally during sample fabrication, methods) can play the role of polariton emitters as well. As shown in Fig. 3a, surface impurities (marked with magenta asterisks) act as point-like emitters that emit circular polariton waves, whereas a surface protrusion (green arrow) act as a line-like emitter. The latter one emit plane waves that do not spread much as they travel. As a result, the amplitude of these waves decays slowly and remains discernible in almost the entire field of view in Fig. 3.

In summary, the imaging data compiled in Figs. 1-3 demonstrate a variety of polariton emitters in hBN of different emission efficiency. Although essentially any topographic feature can act as an emitter, their efficiency is inferior to that of a large metalized tip. Note that the emitters studied in this work may also be employed for scattering polaritons into optical photons detectable by conventional means. The methodology presented in this work may be readily extended to collective modes[2-4] in other vdW materials including graphene[34], transition metal dichalcogenides, black phosphorus, and topological insulators. Our results along with the recent work on grating polaritonic couplers[37] present the initial steps towards developing high-efficiency polariton emitters and detectors for diverse nanophotonics applications.

**Methods**

**Experimental setup**

The nano-imaging experiments described in the main text were performed at UCSD using a commercial s-SNOM (www.neaspec.com). The s-SNOM is based on a tapping-mode AFM illuminated by monochromatic Quantum Cascade Lasers (QCLs) (www.daylightsolutions.com). These lasers cover a frequency range of 900 – 2300 cm$^{-1}$ in the mid-IR. The nanoscale near-field images were registered by a pseudo-heterodyne interferometric detection module with a AFM tapping frequency and amplitude around 280 kHz and 70 nm respectively. In order to obtain the background-free images, the s-SNOM output signal used in this work is the scattering amplitude $s(\omega)$ demodulated at the 3$^{rd}$ harmonics of the tapping frequency.

**Sample fabrication**

Our hBN crystals were exfoliated from bulk samples synthesized with high-pressure techniques and then transferred onto Si wafers with a 300-nm-thick SiO$_2$ layer. The Au patterns were fabricated on the hBN crystals by electron beam lithography.

**Simulation of edge-emitted hyperbolic phonon polariton fringes**

The numerical simulations shown in Fig. 1a (inset) were done within the quasi-static approximation. In this approach the amplitude of the scalar potential $\Phi(x, z)$ in the system is assumed to satisfy the anisotropic Laplace equation: $\partial_x(\varepsilon^x\partial_x\Phi) + \partial_z(\varepsilon^z\partial_z\Phi) = 0$, where the principal values ($\varepsilon^x$, $\varepsilon^z$) of the permittivity tensor are functions of frequency $\omega$. To produce the graphic shown in the inset of Fig. 1a we used the following parameters: (1, 1) for vacuum; (-0.5620 + 0.0678$i$, 2.7782 + 0.0006$i$) for hBN, and (1.4646 + 0.0104$i$, 1.4646 + 0.0104$i$) for SiO$_2$ substrate, which are representative of frequency $\omega$ = 1587 cm$^{-1}$. The 300-nm tall and 2050-nm wide simulation domain surrounded a 100-nm thick hBN slab. At the outer boundary of the domain the $\Phi = -x$ was imposed to model a uniform unit external field in the *x*-direction. This boundary-value problem was solved using MATLAB PDE Toolbox (MATLAB, Inc., Natick, MA; Release 2012b). Shown in Fig. 1a (inset) are the results for the absolute value of the electric field $|E| = [(\partial_x\Phi)^2 + (\partial_z\Phi)^2]^{1/2}$. As one can see, near the corners of the hBN slab, the calculated field distribution exhibits a structure of internal criss-crossing rays, which produce "hot lines" on the surface of the slab. Further away from the edge, the calculated field distribution morphs into a gradually decaying sinusoidal wave. The internal polariton rays and the "hot lines" they produce have the same interpretation as the "hot rings" predicted in simulations and subsequently observed on sidewalls of hBN nanocones[15] and on the top surfaces of hBN slabs deposited on metallic disks[19]. Such high-intensity lines are the beats produced by coherent superpositions of multiple guided waves. As detailed in our previous work,[15,19,21] the guided waves of a slab are discrete eigenmodes characterized by the in-plane momenta $k_l = (\pi l + \chi) \tan\theta / d$, where $l = 0, 1, 2, \ldots$ is the mode number, $\chi \sim 1$ is the boundary condition dependent phase shift, and *d* is the slab thickness. Accordingly, the beats pattern has the periodicity $2d \tan\theta \sim \lambda_P / 4$ while the parameter $\tan\theta = i\sqrt{\varepsilon^z}/\sqrt{\varepsilon^x}$ has the meaning of the slope of the polariton rays with respect to the *z*-axis. Unfortunately, the region where the beats exist is very narrow and we could not probe it with our experimental resolution. On the other hand, the sinusoidal standing waves we have imaged can be identified with the principal $l = 0$ guided waves of wavelength $\lambda_P \equiv 2\pi / k_0$.

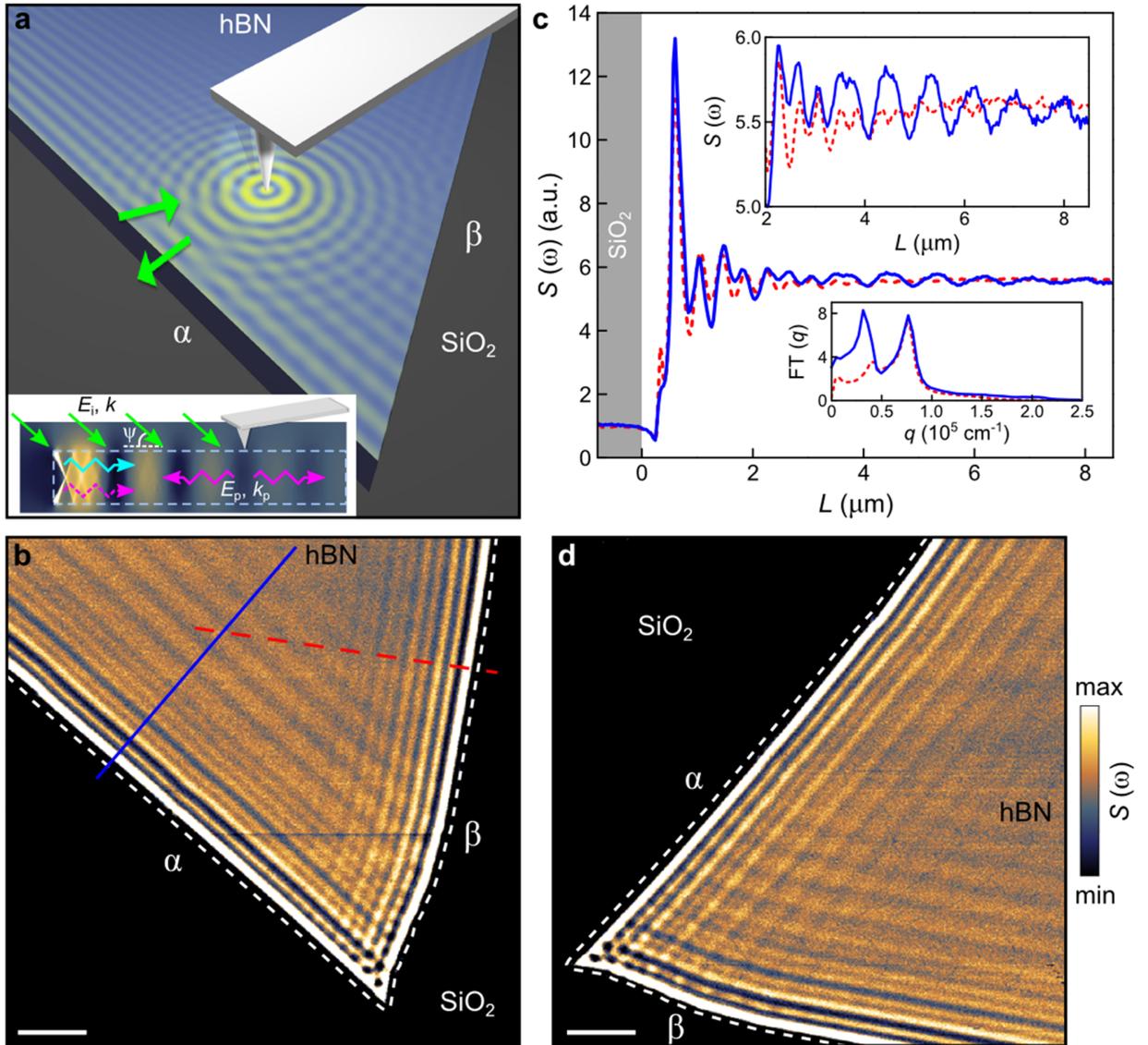

**Figure 1.** Polaritons emitted by the s-SNOM tip and hBN edges. **a**, The experimental schematic. The s-SNOM tip and a tapered hBN crystal are illuminated by the weakly focused IR beam from a Quantum Cascade Laser (QCL). We collect the back-scattered near-field signal (green arrow). Inset, the cross-section of the tip-emitted (magenta arrows) and edge-emitted (cyan arrow) polaritons registered by the tip. Color map, the simulation of the edge-emitted polariton fringes (see Methods). **b**, The near-field amplitude image of the hBN crystal in (**a**). **c,** Polariton line profiles taken perpendicular to the α edge (blue solid line) and β edge (red dashed line) in (**b**). Insets: top, detailed view of the line profiles

when *L* > 2μm. Bottom, the Fourier Transform spectra of the line profiles. **d**, Near-field amplitude image of the same hBN crystal after a clockwise rotation of π/2 from (**b**). White dashed lines track the hBN edges. The hBN thickness: 117 nm. IR frequency: ω = 1530cm$^{-1}$. Scale bar: 2μm.

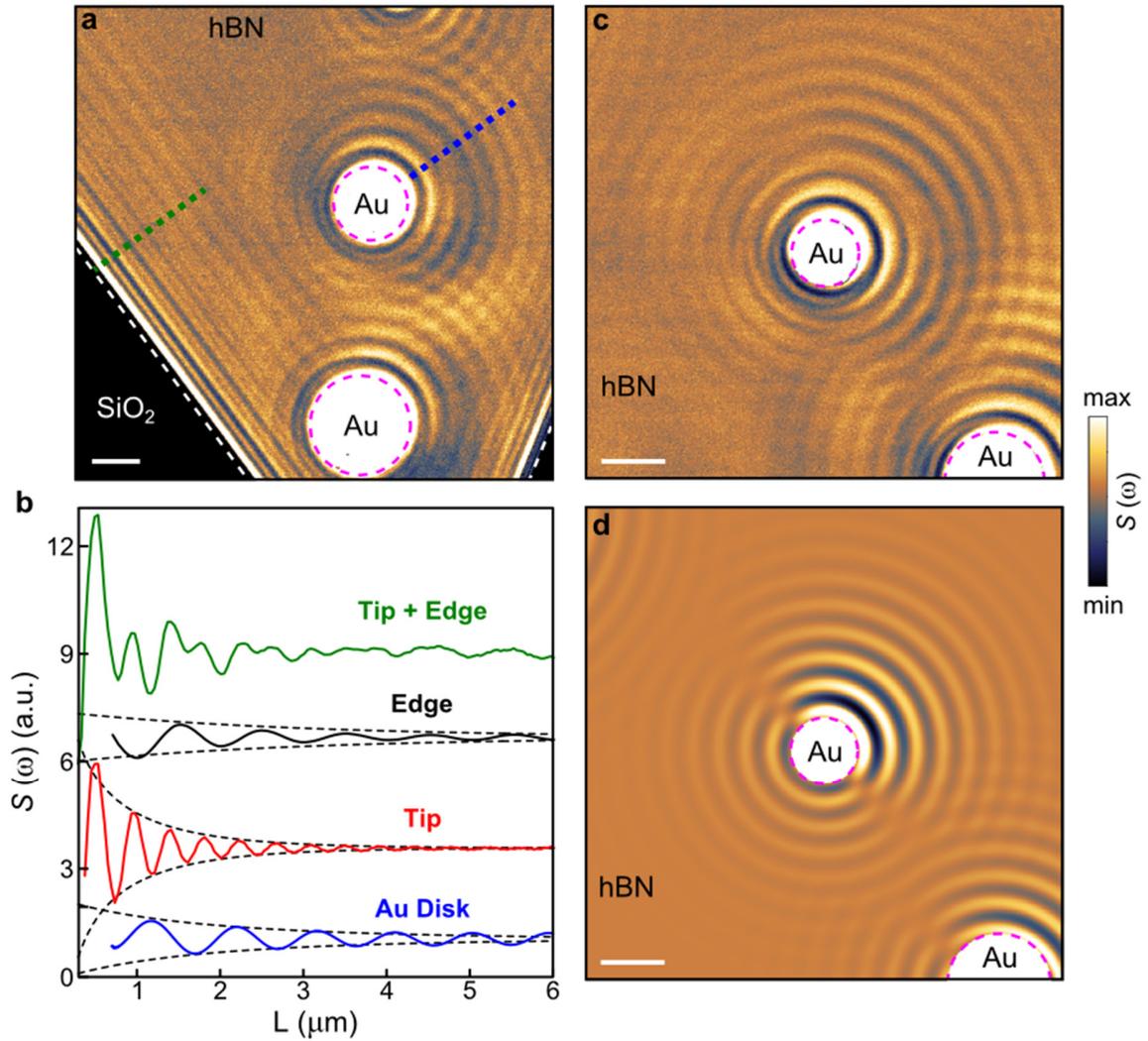

**Figure 2.** Polaritons emission efficiency. **a**, Near-field amplitude image of the hBN crystal with artificially fabricated Au disks on top. **b**, Profiles of polariton fringes as a function of the distance L from the edge of an emitter. The green and blue curves are extracted by averaging a series of linear cuts near the dotted lines in (**a**). The black and red curves are obtained by the FT analysis of the green curve. The dashed curves are the fits to Eq. (1). **c,** Near-field amplitude image of Au disks in the interior of the hBN crystal. **d**, The simulation of s-SNOM image in (**c**). White and magenta dashed lines track the edges of the hBN and Au disks. The hBN thickness: 117 nm. IR frequency: $\omega = 1530 cm^{-1}$. Scale bar: 2μm.

**Table 1. Relative scattering cross-sections and/or scattering lengths for three types of emitters studied.**

| Scattering parameter | s-SNOM tip | hBN edge | Au disk $r_{disk} = 1.5\mu m$ |
|---|---|---|---|
| $l$ | | 0.56 | 1.4 |
| $A$ | 61 | | 14 |

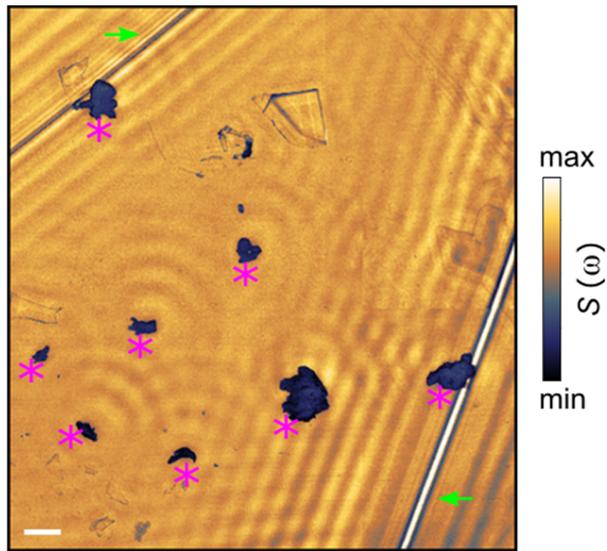

**Figure 3.** Near-field amplitude image of the hBN crystal with impurities (marked with magenta asterisks) and protrusion (green arrow) defect on the surface. The hBN thickness: 253 nm. IR frequency: $\omega = 1550 cm^{-1}$. Scale bar: 2μm.

AUTHOR INFORMATION

**Corresponding Author**

* Email: db3056@columbia.edu (D.N.B.)

ACKNOWLEDGMENT

This research is supported by the ONR under Grant No. N00014-15-1-2671 and the AFOSR Grant No. FA9550-15-1-0478. D. N. B. is an investigator in Quantum Materials funded by the Gordon and Betty Moore Foundation's EPiQS Initiative through Grant No. GBMF4533. P.J-H acknowledges support from AFOSR grant number FA9550-11-1-0225 and the Packard Fellowship program.

ABBREVIATIONS

s-SNOM, scattering-type scanning near-field optical microscopy; hBN, hexagonal boron nitride; vdW, van der Waals; $HP^2$, hyperbolic phonon polariton; IR, infrared; AFM, atomic force microscope; FT, Fourier Transform.